\documentclass[runningheads]{llncs}
\usepackage[T1]{fontenc}
\usepackage{microtype,mathtools,tikz,thm-restate,enumitem}

\usepackage{orcidlink}
\renewcommand{\orcidID}[1]{\unskip\,\orcidlink{#1}}
\hypersetup{hidelinks}

\newcommand{\NP}{\text{NP}}
\newcommand{\cD}{\mathcal{D}}
\newcommand{\cO}{\mathcal{O}}
\newcommand{\cS}{\mathcal{S}}
\newcommand{\clean}{conchort}
\newcommand{\dist}{\delta}

\def\myscale{0.75}
\tikzstyle{vertex}=[draw,circle,fill=black,inner sep=1pt]
\tikzstyle{ivertex}=[draw,circle,inner sep=.8pt,fill=white]
\tikzstyle{edges}=[draw=black!66, line join=round]
\newcommand{\drawGraphlet}[3]{
  \foreach \vertex/\pos/\labelSide in {#1} {
    \coordinate (v-\vertex) at (\pos);}

  \foreach \vertexA/\vertexB in {#2} {
    \draw[edges] (v-\vertexA) -- (v-\vertexB);}
    
  \foreach \vertex/\pos\labelSide in {#1} {
    \node[vertex] at (v-\vertex) {};
    \ifnum #3>0 
      \node[\labelSide] at (v-\vertex) {$\vertex$};
    \fi}
}

\begin{document}

\title{Linear-Time Optimal Leaf Roots\\ beyond 3-Leaf Powers}
\titlerunning{Linear-Time Optimal Leaf Roots beyond 3-Leaf Powers}

\renewcommand{\thefootnote}{\fnsymbol{footnote}}

\author{Daniela Ceban\orcidID{0009-0002-1818-5456} \and Christian Rosenke\orcidID{0009-0003-8222-8366}{\footnotesize\textsuperscript{$\star$}}}
\authorrunning{D. Ceban and C. Rosenke}
\institute{Institut f\"ur Informatik, Universit\"at Rostock, Rostock, Germany \\
  \email{daniela.ceban@uni-rostock.de}\\
  \email{christian.rosenke@uni-rostock.de}}
\maketitle

\footnotetext[1]{Corresponding author.}
%
\begin{abstract}
A $k$-leaf power for an integer $k \geq 2$ is a graph $G$ that permits a $k$-leaf root, a tree $T$ with leaf set $V(G)$ such that, for all distinct vertices $x,y \in V(G)$, their distance in $T$ is at most $k$ if and only if $xy$ is an edge of $G$.
Constructing such trees is motivated by the reconstruction of phylogenetic trees from similarity data.

For every fixed $k$, Lafond~[ACM Trans. Algorithms 2023] showed that $k$-leaf powers can be recognized in polynomial time.
Recently, Dupr\'e la Tour, Lafond, and Ndiaye~[SODA 2026] proved that the problem becomes $\NP$-complete when $k$ is part of the input and that deciding whether a graph is a $k$-leaf power for some $k$ is $\NP$-complete as well.

Le and Rosenke~[J. Graph Algorithms Appl. 2024] proposed a different algorithmic perspective by studying \emph{optimal leaf roots} of graphs $G$, that is, the $k$-leaf roots of $G$ with minimum possible $k$.
This smallest $k$-value is called the \emph{leaf rank} of $G$.
They showed that optimal leaf roots of chordal cographs, a class of unbounded leaf rank, can be constructed in linear time.
However, chordal cographs do not even contain all $3$-leaf powers.

This paper extends linear-time optimal-leaf-root construction to three natural superclasses of the $3$-leaf powers, the $(\text{bull},\text{gem})$-free, the $(\text{bull},\text{dart})$-free, and the $(\text{dart},\text{gem})$-free chordal graphs.
Except for the latter case, which follows from known results, our methods are based on structural decompositions of bull-free chordal graphs.
In particular, the relevant $(\text{bull},\text{gem})$-free graphs decompose into chordal cographs attached to a tree, while the $(\text{bull},\text{dart})$-free case reduces to trees and powers of paths.

\keywords{$k$-leaf power \and $k$-leaf root \and Optimal $k$-leaf root \and Chordal cographs \and $3$-leaf powers \and Bull-free leaf powers}
\end{abstract}

\section{Introduction}
\label{sec:introduction}

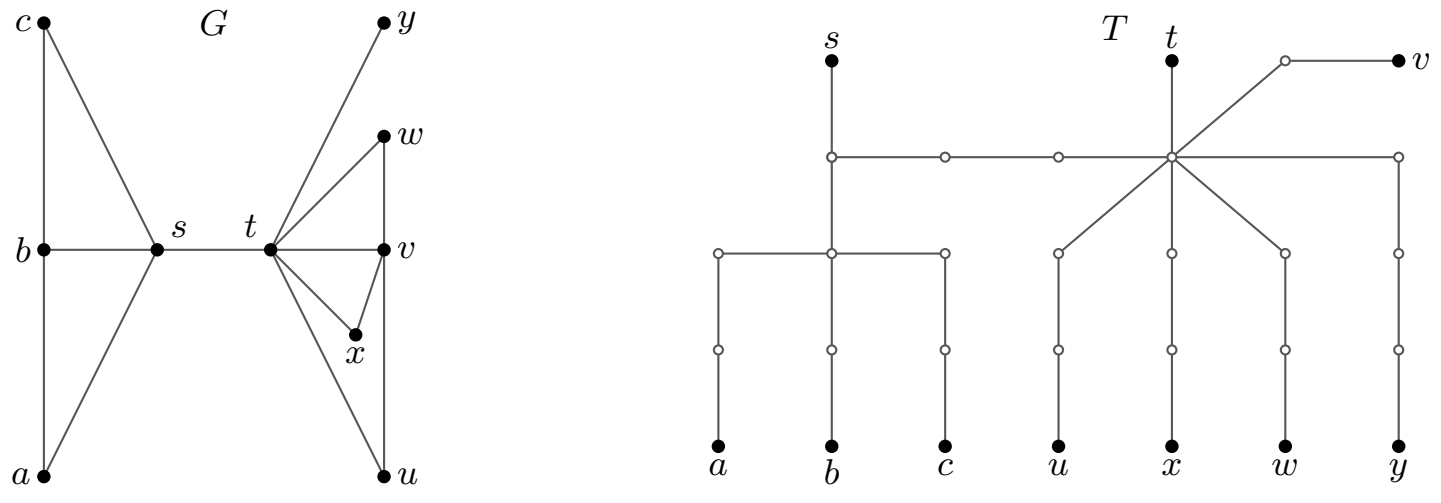
\begin{figure}[!ht]\centerline{
  \begin{tikzpicture}[scale = \myscale, every node/.style={scale=\myscale}]
    	  \node at (1.5,4) {$G$};
        \drawGraphlet{a/{0,0}/left, b/{0,2}/left, c/{0,4}/left, s/{1,2}/above right, u/{3,0}/right, v/{3,2}/right, w/{3,3}/right, x/{2.75,1.25}/below, y/{3,4}/right, t/{2,2}/above left}{
        a/b,b/c,a/s,b/s,c/s,
        s/t,
        u/v,v/w,v/x,u/t,v/t,w/t,x/t,y/t}{1}
  \end{tikzpicture}
  \hspace*{1.5cm}
  \begin{tikzpicture}[scale = \myscale, yscale=0.85, every node/.style={scale=\myscale}]
    \node at (2.5,-0.66) {$T$};
    \draw[edges] (-1,-5) coordinate (v-a) -- ++(0,1) node[ivertex] {} -- ++(0,1) node[ivertex] {} -- ++(1,0);
    \draw[edges] (1,-5) coordinate (v-c) -- ++(0,1) node[ivertex] {} -- ++(0,1) node[ivertex] {} -- ++(-1,0);
    \draw[edges] (0,-5) coordinate (v-b) -- ++(0,1) node[ivertex] {} -- ++(0,1) node[ivertex] {} -- ++(0,1) node[ivertex] {};
    \draw[edges] (2,-5) coordinate (v-u) -- ++(0,1) node[ivertex] {} -- ++(0,1) node[ivertex] {} -- ++(1,1);
    \draw[edges] (3,-5) coordinate (v-x) -- ++(0,1) node[ivertex] {} -- ++(0,1) node[ivertex] {} -- ++(0,1);
    \draw[edges] (4,-5) coordinate (v-w) -- ++(0,1) node[ivertex] {} -- ++(0,1) node[ivertex] {} -- ++(-1,1);
    \draw[edges] (5,-5) coordinate (v-y) -- ++(0,1) node[ivertex] {} -- ++(0,1) node[ivertex] {} -- ++(0,1) node[ivertex] {} -- ++(-2,0) -- ++(1,1) node[ivertex] {} -- ++(1,0) coordinate (v-v);
    \draw[edges] (0,-1) coordinate (v-s) -- ++(0,-1) node[ivertex] {} -- ++(1,0) node[ivertex] {} -- ++(1,0) node[ivertex] {} -- ++(1,0) node[ivertex] {} -- ++(0,1) coordinate (v-t);
    \foreach \vertex/\labelSide in
      {a/below, b/below, c/below, s/above, t/above, u/below, v/right, w/below, x/below, y/below}
    {
      \node[vertex] at (v-\vertex) {};
      \node[\labelSide] at (v-\vertex) {$\vertex$};}
  \end{tikzpicture}}
  \caption{A graph $G$ and a $5$-leaf root $T$ of $G$}
  \label{fig:leaf_root_example}
\end{figure}

Introduced by Nishimura, Ragde, and Thilikos~\cite{NishimuraRT02}, leaf powers model the phylogeny reconstruction problem from computational biology:
given a graph $G$ whose vertices represent species and whose edges represent interspecies similarities, reconstruct an evolutionary tree $T$ explaining these similarities.
Formally, a \emph{$k$-leaf root} of $G$, for an integer $k\geq 2$, is a tree $T$ with leaf set $V(G)$ such that distinct species $x,y\in V(G)$ have distance at most $k$ in $T$ if and only if $xy\in E(G)$.
If such a tree exists, $G$ is a $k$-leaf power.
Figure~\ref{fig:leaf_root_example} illustrates a graph together with a $5$-leaf root.

Algorithmic research on $k$-leaf roots has primarily focused on the corresponding \emph{$k$-leaf power recognition problem}: given $G$ and $k$, decide if $G$ has a $k$-leaf root.
This computational problem has been studied extensively since the introduction of the notion.
Among the first milestones are structural characterizations and linear-time recognition algorithms for $3$- and $4$-leaf powers~\cite{BrandstadtL06,BrandstadtLS08} and recognition algorithms for $5$- and $6$-leaf powers~\cite{ChangK07,Ducoffe19}.
For a more comprehensive overview, we refer to the chapter on leaf powers in \emph{Topics in Algorithmic Graph Theory}~\cite{RosenkeLB21}.
However, two important developments after the publication of that chapter have changed the algorithmic landscape considerably.
Lafond~\cite{Lafond23} showed that $k$-leaf power recognition can be solved in $\cO(n^{g(k)})$ time, where $g$ depends only on $k$.
Hence, the problem is in polynomial time for fixed $k$.
When $k$ is part of the input, even with unary encoding, Dupr\'e la Tour, Lafond and Ndiaye~\cite{DupreLN26} recently proved that recognition is $\NP$-complete.
They also show the $\NP$-completeness of deciding for a given graph $G$ if there is an integer $k$ such that $G$ is a $k$-leaf power.

In view of the severe dependency of leaf root construction on the parameter $k$, Le and Rosenke~\cite{LeR24} pursued the following optimization problem:
given a graph $G$, the \textsc{Optimal Leaf Root} problem (OLR) asks for a $\kappa$-leaf root $T$ of $G$ such that $\kappa\leq k$ for every $k$-leaf root of $G$, or answers \emph{No}, if $G$ is not a leaf power.
The integer $\kappa$ is also referred to as the \emph{leaf rank} of $G$ in the literature~\cite{BrandstadtHMW10,Hogemo24}.
From the motivating viewpoint, minimizing $k$ is natural because it delivers a phylogenetic tree that minimizes the maximum number of evolutionary events that may lie between similar species.
The main result in~\cite{LeR24} is a linear-time algorithm that solves OLR for chordal cographs, which are exactly the $P_4$-free chordal graphs, a graph class of unbounded leaf rank also known as the trivially perfect graphs.

For example, the $5$-leaf root in Figure~\ref{fig:leaf_root_example} is optimal for the given graph $G$, which, accordingly, has leaf rank $5$.
However, $G$ contains an induced $P_4$.

Chordal cographs do not even contain all $3$-leaf powers, which usefully can be described as follows:
A connected graph is a $3$-leaf power if and only if it is a $(\text{bull},\text{dart},\text{gem})$-free chordal graph~\cite{DomGHN06} if and only if it can be obtained from a tree by substituting non-empty cliques into its vertices~\cite{BrandstadtL06}.
In contrast to chordal cographs, $3$-leaf powers may contain $P_4$, but they exclude, among others, bull and gem, two natural extensions of $P_4$.
They additionally exclude the dart, which is itself a chordal cograph and, thus, explains why chordal cographs are not a subclass of $3$-leaf powers.
Figure~\ref{fig:small_graphs} shows the three graphs bull, dart, and gem.

In this paper, we relax the forbidden subgraph characterization of $3$-leaf powers and study the complexity of OLR on the three classes of chordal graphs obtained by forbidding only two graphs:
the $(\text{bull},\text{dart})$-free, the $(\text{bull},\text{gem})$-free, and the $(\text{dart},\text{gem})$-free chordal graphs.
Naturally, each of them generalizes $3$-leaf powers, but only the $(\text{bull},\text{gem})$-free chordal graphs also contain the chordal cographs.
Our main algorithmic result is the following.
\begin{theorem}\label{thm:main}
  Let $G$ be a chordal graph on $n$ vertices and $m$ edges.
  If $G$ is $(\text{bull},\text{gem})$-free, $(\text{bull},\text{dart})$-free, or $(\text{dart},\text{gem})$-free, then an optimal leaf root of $G$ can be computed in $\cO(n+m)$ time.
\end{theorem}
Before we can elaborate on the ideas behind this result, we remind the reader of the well-established fact in this area that a graph $G$ is a $k$-leaf power if and only if every connected component of $G$, after collapsing each true twin class to one representative, is a $k$-leaf power.
Therefore, reducing recognition or leaf root construction to \emph{con}nected \emph{chor}dal graphs without \emph{t}rue twins is common practice in the given research field.
To avoid repeatedly defining this condition in the remainder of the paper, we call these graphs \emph{\clean}.

It turns out that the $(\text{dart},\text{gem})$-free case of Theorem~\ref{thm:main} follows from known results.
In fact, a chordal graph is $(\text{dart},\text{gem})$-free if and only if it is obtained from a block graph by substituting cliques into its vertices~\cite{BrandstadtL09}.
That $k$-leaf powers are closed under clique substitution, being the reverse operation to true twin reduction, is folklore.
Thus, it suffices to consider just the underlying block graphs.
Every block graph is a $4$-leaf power~\cite{BrandstadtW07} and, hence, has leaf rank $2$, $3$, or $4$.
Since $2$-leaf powers are trivial and for both $3$- and $4$-leaf powers, recognition and root construction are in linear time~\cite{BrandstadtL06,BrandstadtLS08}, OLR can be solved in linear time, here.

For the other two classes, we develop the following main structural tool.
\begin{restatable}{theorem}{bullfreedecomposition}\label{thm:bull_free_decomposition}
  A graph $G$ is a bull-free {\clean} graph if and only if one of the following is true:
  \begin{enumerate}[label=(\roman*)]
    \item\label{thm:bull_free_decomposition:One} $G$ is a single vertex or a $2$-connected bull-free {\clean} graph.
    \item\label{thm:bull_free_decomposition:Two} $G$ has a universal cut vertex $u$, and every connected component of $G-u$ is a smaller bull-free {\clean} graph.
    \item\label{thm:bull_free_decomposition:Three} There is a tree $H$ on at least four vertices where distinct vertices in the leaf set $L$ have different neighbors in $H$ and $G$ results from $H$ by substituting every leaf $x \in L$ with a bull-free chordal graph $G_x$ without true twins.
  \end{enumerate}
\end{restatable}
Theorem~\ref{thm:bull_free_decomposition} reduces the remaining cases to the structure of their $2$-connected graphs and, for the $1$-separable members, to the structure of graphs attached to the leaves of the tree $H$ underlying the decomposition.
For $(\text{bull},\text{dart})$-free {\clean} graphs, both parts of the theorem become particularly rigid.
We show that the $1$-separable cases are precisely the trees on at least four vertices, while every $2$-connected graph is proper interval, here.
The former condition represents the $3$-leaf powers after true twin reduction and the latter is much stronger than it first appears.
More precisely, every connected bull-free proper interval graph $G$ without true twins is isomorphic to the path power $P_n^{\omega(G)-1}$.
In both situations, the optimal leaf root can be obtained with little additional effort.

The $(\text{bull},\text{gem})$-free setting represents the main algorithmic difficulty, although the $2$-connected case cannot occur.
The $1$-separable graphs, with the additional exclusion of gem, make induced $P_4$ in the substituted graphs $G_x$ impossible and, thus, reduce them to chordal cographs. 
This enables the combination of locally optimal leaf roots for suitably augmented versions of the graphs $G_x$, obtained by the algorithm from~\cite{LeR24}, with the core of tree $H$ without losing global optimality.
Together, the three ways of generalizing $3$-leaf powers considerably extend the graph classes for which OLR can be solved in linear time.

The example graph $G$ in Figure~\ref{fig:leaf_root_example} is $(\text{bull},\text{gem})$-free {\clean} and, in fact, the given $5$-leaf root $T$ is constructed with the new algorithm.

The remainder of the paper is organized as follows.
After Section~\ref{sec:preliminaries} introduces our notation, Section~\ref{sec:structure} develops the structural results for the considered subclasses of chordal graphs, while Section~\ref{sec:leaf-roots} derives the corresponding optimal leaf root construction and proves Theorem~\ref{thm:main}.
All proofs except the forward direction of Theorem~\ref{thm:bull_free_decomposition} are deferred to the appendix.

\section{Preliminaries}
\label{sec:preliminaries}

All graphs considered in this paper are finite, simple, and undirected.
For (vertex) sets $X$ and $Y$, we use $X + Y$ and $X - Y$ for their union and difference.
Moreover, $X+x$ is used as a shorthand for $X + \{x\}$, if this is clear from the context.

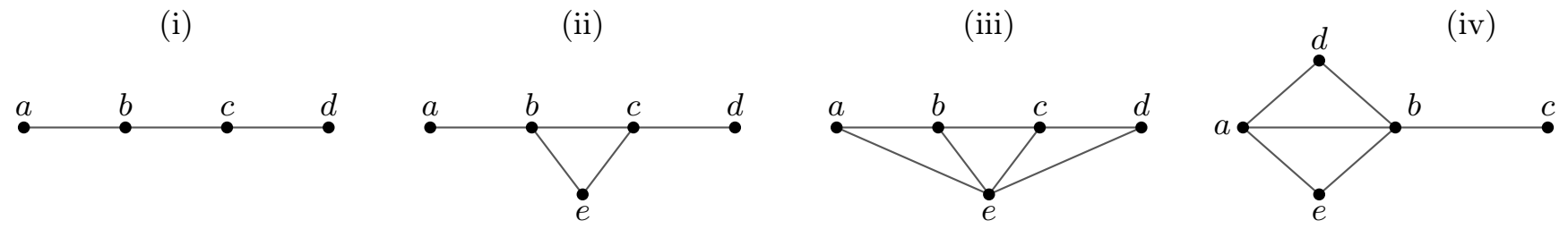
\begin{figure}[!ht]
  \centerline{
  \begin{tikzpicture}[scale = \myscale, every node/.style={scale=\myscale}]
	\begin{scope}[xshift=0cm]
    	\drawGraphlet{a/{0,0}/above,b/{1,0}/above,c/{2,0}/above,d/{3,0}/above}{a/b,b/c,c/d}{1}
      \node at (1.5,1) {(i)};
	\end{scope}
	\begin{scope}[xshift=4cm]
    	\drawGraphlet{a/{0,0}/above,b/{1,0}/above,e/{1.5,-0.66}/below,c/{2,0}/above,d/{3,0}/above}{a/b,b/c,c/d,b/e,c/e}{1}
      \node at (1.5,1) {(ii)};
	\end{scope}
  \begin{scope}[xshift=8cm]
    	\drawGraphlet{a/{0,0}/above,b/{1,0}/above,e/{1.5,-0.66}/below,c/{2,0}/above,d/{3,0}/above}{a/b,b/c,c/d,a/e,b/e,c/e,d/e}{1}
      \node at (1.5,1) {(iii)};
	\end{scope}
  \begin{scope}[xshift=12cm]
    	\drawGraphlet{a/{0,0}/left,b/{1.5,0}/above right,c/{3,0}/above,d/{0.75,0.66}/above,e/{0.75,-0.66}/below}{a/b,b/c,a/d,b/d,a/e,b/e}{1}
      \node at (2.25,1) {(iv)};
	\end{scope}
  \end{tikzpicture}}
  \caption{(i) $P_4$, (ii, iii) bull and gem, respectively, with $P_4$ on $a,b,c,d$ and \emph{tip} $e$,\\ (iv) dart with $P_3$ on $a,b,c$ and \emph{wing tips} $d,e$}
  \label{fig:small_graphs}
\end{figure}

The small graphs used throughout the paper are shown in Figure~\ref{fig:small_graphs}.

For a graph $G$, vertex and edge set are denoted as $V(G)$ and $E(G)$.
Let $x \in V(G)$.
Then $N_G(x) = \{y \mid xy \in E(G)\}$ is the \emph{open neighborhood} of $x$ and $N_G[v]=N_G(v)+v$ the closed one.
The \emph{degree} of $x$ is defined as $|N_G(x)|$.
A vertex $y$ is a \emph{true twin} of $x$, if $N_G[x]=N_G[y]$.
If $N[x] = V(G)$ then $x$ is \emph{universal}.

For $X\subseteq V(G)$, the \emph{induced subgraph} $G[X]$ has vertices $X$ and edges $\{xy \in E(G) \mid x,y \in X\}$.
We use $G-X$ for $G[V(G)-X]$, if $X=\{x\}$ then simply $G-x$, and $H+x$ for $G[V(H)+x]$, if $H$ is induced subgraph of $G$ with $x \in V(G)-V(H)$.
Graphs $G$ and $H$ are \emph{isomorphic}, if there is an adjacency-preserving bijection between their vertex sets.
A graph is $H$-free if no induced subgraph is isomorphic to $H$ and $(H_1,\dots,H_t)$-free if it is $H_i$-free for every $i\in \{1,\dots,t\}$.
A \emph{clique} is a subgraph of mutually adjacent vertices.
The maximum cardinality of a clique in $G$ is denoted by $\omega(G)$.
A clique on three vertices is a \emph{triangle}.

For $x,y\in V(G)$, an $x,y$-\emph{path} is a sequence $x=v_1,\dots,v_t=y$ of distinct vertices such that $v_iv_{i+1}\in E(G)$ for every $i \in \{1,\dots,t-1\}$.
Its \emph{length} is $t-1$, and it is \emph{shortest}, if no $x,y$-path has length less than $t-1$.
The distance $\dist_G(x,y)$ between $x$ and $y$ is the length of a shortest $x,y$-path in $G$.
The path is \emph{induced}, if $v_iv_j \notin E(G)$ whenever $|i-j|>1$.
Clearly, a shortest $x,y$-path is induced.
A \emph{cycle} is a sequence $v_1,\dots,v_t,v_1$ of $t\geq 3$ distinct vertices $v_1,\dots,v_t$ such that consecutive vertices, including $v_tv_1$, are adjacent.
Its length is $t$, and it is induced if no two non-consecutive vertices of the cycle are adjacent.
Throughout the paper, $P_t$ refers to an induced path and $C_t$ to an induced cycle on $t$ vertices.

A graph is \emph{chordal} if it contains no induced cycle of length at least four and a \emph{cograph} if it contains no induced $P_4$.
Thus, the chordal cographs are exactly the chordal $P_4$-free graphs and are also known as \emph{trivially perfect graphs}.

A graph is \emph{connected} if every pair $x, y$ of vertices are joined by an $x,y$-path, otherwise, \emph{disconnected}.
To avoid repeatedly referring to connected chordal graphs without true twins, we call such graphs \emph{\clean} throughout the paper.

Let $G$ be a connected graph.
If $G$ is cycle-free it is a \emph{tree}.
A vertex $c$ of $G$ is a \emph{cut vertex} if $G-c$ is disconnected.
We call $G$ \emph{$1$-separable} if it has a cut vertex.
If $G$ does not contain a cut vertex and $|V(G)|\geq 3$ then $G$ is \emph{$2$-connected}.
An edge is a \emph{bridge} if removing it from $E(G)$ makes $G$ disconnected.
A \emph{block} of $G$ is either a maximal $2$-connected subgraph or a bridge including its endpoints.
Clearly, every edge of $G$ belongs to exactly one block.
The \emph{block-cut tree} $\text{BCT}$ of $G$ is defined on the bipartite node set of the blocks and the cut vertices by making block-node and cut-node adjacent exactly when the cut vertex belongs to the block.

For a graph $G$ and an integer $d\geq 1$, the $d$th power $G^d$ is the graph on $V(G)$ where distinct vertices are adjacent exactly if their distance in $G$ is at most $d$.
In particular, $P_t^d$ is the $d$th power of a $t$-vertex path.
A graph $G$ that consists of a path $P$ with pendant edges to a subset of its vertices is a tree called a \emph{caterpillar}.
The path $P$ is also called the \emph{spine} of $G$.

Let $G$ and $H$ be vertex-disjoint and $x\in V(H)$.
To \emph{substitute $G$ for $x$ in $H$} means to remove $x$, add $G$, and make all vertices of $G$ adjacent to $N_H(x)$, not adding further edges between $G$ and $H-x$.
We use the same terminology when multiple graphs $G_x$ are substituted simultaneously for several vertices $x$ of $H$.

A graph is a \emph{proper interval graph} if it is the intersection graph of a family of intervals on the real line in which no interval properly contains another one.
Equivalently, a proper interval graph admits a so-called \emph{proper interval ordering} $<_G$ of $V(G)$ where $u <_G v <_G w$ and $uw\in E(G)$ implies $uv,vw\in E(G)$~\cite{CorneilKNOS95}.

A degree-one vertex $x$ of a tree is a \emph{leaf} and the edge incident to $x$ is a \emph{pendant edge}.
For $k\geq 2$, a graph $G$ is a \emph{$k$-leaf power} if there is a tree $T$ with leaf set $V(G)$ such that, for all distinct $x,y\in V(G)$, $xy \in E(G)$ if and only if $\dist_T(x,y)\leq k$.
Then $T$ is called a \emph{$k$-leaf root} of $G$.
A $k$-leaf root is \emph{optimal} if $k\leq k'$ for every $k'$-leaf root of $G$.
The minimum possible value of $k$ is called the \emph{leaf rank} of $G$.

As in~\cite{LeR24}, the parity $\pi(k)=k\bmod 2$ of parameter $k$ plays an important role in optimal leaf root construction.
For $p \in \{0,1\}$, a $k$-leaf root $T$ of $G$ is \emph{optimal of parity $p$} if $\pi(k)=p$ and $k\leq k'$ for every $k'$-leaf root of $G$ with $\pi(k')=p$.

Like in~\cite{LeR24}, we frequently represent $k$-leaf roots \emph{compressed}, where edges $xy$ are weighted with a positive integer $\ell$ to represent $x,y$-paths of length $\ell$ with degree-two internal vertices.
Distances in such a tree are the edge-weight-sums along corresponding paths.

We also reuse two operations from~\cite{LeR24}.
For a connected chordal cograph $G$ without true twins, $(T,k)=\rho(G,p)$ denotes the linear-time computation of a compressed $k$-leaf root $T$ for $G$ that is optimal of parity $p$.
Moreover, for a $k$-leaf root $T$ of $G$ and an integer $\delta\geq 0$, the $(k+2\delta)$-leaf root $\eta(T,\delta)$ results from replacing all pendant edges of $T$ with paths of length $\delta+1$.

\section{The Structure of Bull-Free Conchort Graphs}
\label{sec:structure}

Both remaining generalizations of $3$-leaf powers are subclasses of bull-free {\clean} graphs.
A meaningful way to attack OLR on both of them is to uncover tree-like substructures underlying chordal graphs that forbid bull as induced subgraph.
The hope in that is to find a general method of fitting leaf roots along the branches of such a substructure.
However, at the time of writing this article, it is even unknown whether all bull-free {\clean} graphs are leaf powers.


\subsection{Decompositions of $1$-separable Bull-Free Conchort Graphs}
\label{sec:bull-free-structure}
In a bull-free graph, studying the cut vertices $c$ is a very natural starting point.
The bull itself contains two cut vertices and, therefore, forbidding induced bulls imposes strong restrictions on how different blocks can intersect in $c$.
In fact, we can easily see that $c$ has to be universal in each of its blocks.
\begin{restatable}{lemma}{cutvertexblock}\label{lem:bull-free_structure:cutvertices}
  If $G$ is a bull-free {\clean} graph, $c$ is a cut vertex of $G$, and $B$ is a block of $G$ that contains $c$, then $c$ is adjacent to all vertices in $V(B)- c$.
\end{restatable}
The reason that a cut vertex $c$ sees everything in a block $B$ is that, otherwise, there would be a vertex $u \in V(B)-N_B(c)$ that shares a neighbor $v$ with $c$.
Together with any neighbor $x \in N_G(c) - V(B)$ of $c$ outside the block, these four vertices $u,v,c,x$ would induce a $P_4$ in $G$.
Moreover, that $B$ is a $2$-connected component would imply another vertex $w \in V(B)$ with $u,v,c \in N_G(w)$ and $wx \notin E(G)$.
Now, since none of the vertices $u,c,x$ distinguishes $v$ and $w$ and because $G$ is true twin-free, there is a vertex $y$ that is adjacent to exactly one of them, say $v$.
Browsing through all possible neighborhood relations between $y$ and the vertices $u,c,x$ delivers a forbidden induced subgraph in every constellation.

The local universality of cut vertices has a tremendous effect on the connectivity of the blocks within $G$.
In fact, we can conclude that every block can host at most one cut vertex unless it is a bridge.
\begin{restatable}{lemma}{twocutvertices}\label{lem:bull-free_structure:two_cutvertices}
  If $G$ is a bull-free {\clean} graph and $B$ is a block of $G$, then $B$ is an edge or contains at most one cut vertex of $G$.
\end{restatable}
The lemma uses that two cut vertices $u$ and $v$ in a block $B$, which has at least one more vertex $w$, would first imply the existence of vertices $x \in N_G(u) - V(B)$ and $y \in N_G(v) - V(B)$ and second that $u,v,w$ would be a triangle due to Lemma~\ref{lem:bull-free_structure:cutvertices}.
Since $u$ and $v$ would separate $x$ and $y$ from the rest of $V(B)$, $x$ would see $u$ but neither $v$ nor $w$ and $y$ would see $v$ but not $u$ or $w$.
Hence, $G$ would contain an induced $C_4$, if $xy \in E(G)$, or an induced bull, otherwise.

Recall that it is always the case that the blocks of a connected graph $G$ have tree structure.
This is in particular demonstrated by the block-cut-tree of $G$.
With Lemma~\ref{lem:bull-free_structure:two_cutvertices}, however, the block structure becomes even more restricted for bull-free {\clean} graphs.
In fact, inner blocks, those that share more than one cut vertex with neighboring blocks, are required to be just edges, while the outer blocks with only one cut vertex correspond to the leaves of the block-cut-tree.
This condition is precisely formalized in Theorem~\ref{thm:bull_free_decomposition}, which we subsequently prove.

\begin{proof}[$\Rightarrow$-direction of Theorem~\ref{thm:bull_free_decomposition}]
Let $G$ be a bull-free {\clean} graph.
The idea of the proof is to handle three mutually excluding properties of this graph, each leading to one of the three conditions stated in Theorem~\ref{thm:bull_free_decomposition}.
The first is when $G$ is a single vertex or $2$-connected.
Then $G$ trivially satisfies Condition~\ref{thm:bull_free_decomposition:One} and there is nothing to show.
So, in the remainder of the proof, let $G$ be $1$-separable.

As the second possible property of $G$, we suppose that $G$ has exactly one cut vertex $u$.
Then $G-u$ decomposes into connected components $G_1, \dots, G_t$.
Since $u$ is the only cut vertex of $G$, the graphs $B_1 = G_1+u, \dots, B_t = G_t+u$ are exactly the blocks of $G$.
Hence, every block of $G$ contains $u$.
According to Lemma~\ref{lem:bull-free_structure:cutvertices}, the vertex $u$ is universal in every block of $G$ and, thus, $u$ is universal in $G$.
Each of the graphs $G_1, \dots, G_t$ is a connected induced subgraph of $G$ and is therefore bull-free and chordal.
It remains to show that every connected component of $G-u$ is true twin-free.
So, suppose that $x,y \in V(G_i)$ were true twins in $G_i$ for some $i \in \{1, \dots, t\}$.
Since $u$ is adjacent to every vertex of $G_i$ and neither $x$ nor $y$ has a neighbor in another component of $G-u$, the vertices $x$ and $y$ would also be true twins in $G$, a contradiction.
Therefore, $G_1, \dots, G_t$ are bull-free {\clean} graphs and $G$ satisfies Condition~\ref{thm:bull_free_decomposition:Two}.

Finally, we consider the third potential property of $G$, where $G$ has at least two cut vertices and which will lead to Condition~\ref{thm:bull_free_decomposition:Three}.
While every block of $G$ contains at least one cut vertex, here, Lemma~\ref{lem:bull-free_structure:two_cutvertices} also states that all blocks contain at most two cut vertices.
Moreover, every block with two cut vertices is an edge.
It is therefore meaningful to partition the blocks of $G$ into two classes.
Accordingly, let $\cS$ be the set of blocks that contain a \emph{single} cut vertex, and let $\cD$ be the set of blocks with a \emph{duo} of cut vertices.
Thus, $\cD$ is a possibly empty set of edge-blocks.
Later, the edges in $\cD$ will form the tree $H$ up to pendant edges.

Let $\{c_1,\dots,c_t\}$ be the set of cut vertices that occur as the unique cut vertex of some block in $\cS$.
Notice that this set does not necessarily cover all cut vertices of $G$ because it misses those that only appear in the edge-blocks of $\cD$.
For every cut vertex $c \in \{c_1,\dots,c_t\}$, we let $\cS_c = \{B \in \cS \mid c\in V(B)\}$.
According to Lemma~\ref{lem:bull-free_structure:cutvertices}, the vertex $c$ is universal in every block $B\in\cS_c$.
Furthermore, if $B,B'\in\cS_c$ are distinct, then $V(B) \cap V(B') = c$, and there is no edge between $B-c$ and $B'-c$.

For every $c \in \{c_1,\dots,c_t\}$, we now define
\[
G'_c= G\left[\bigcup_{B\in\cS_c}\bigl(V(B)-\{c\}\bigr)\right].
\]
Hence, $G'_c$ is the not necessarily connected induced subgraph of $G-c$ that covers all single-cut-vertex-blocks incident to $c$, except for the vertex $c$ itself.
Moreover, no vertex of $G'_c$ has a neighbor in $V(G) - V(G'_c)- c$.
Later, every graph $G'_c$ will be renamed to $G_x$ for a certain leaf $x$ of $H$ and then substituted into $x$.

As $G'_c$ is an induced subgraph of $G$, it is bull-free and chordal.
Suppose that $y,z\in V(G'_c)$ were true twins in $G'_c$.
Then $y$ and $z$ would be adjacent and, hence, they would belong to $B-c$ for the same block $B \in \cS_c$.
Since $c$ is adjacent to both $y$ and $z$, and neither vertex has a neighbor outside $V(G'_c) + c$, the vertices $y$ and $z$ would be true twins in $G$, a contradiction.
Hence, every $G'_c$ is a bull-free chordal graph without true twins.

To prepare the creation of the tree $H$, we now define
\[
h = G - \bigcup_{i=1}^t V(G'_{c_i}).
\]
Every non-cut vertex of $G$ belongs to a single-block and is therefore contained in one of the graphs $G'_{c_i}$.
Consequently, the vertex set of $h$ consists precisely of all cut vertices of $G$, and its edges are precisely the edge-blocks from $\cD$.

We want to show that $h$ is a tree and, first, argue that $h$ is connected.
Consider any pair of distinct cut vertices $c$ and $c'$ of $G$.
In the connected graph $G$, let $c = v_1, \dots, v_s = c'$ be a shortest $c,c'$-path.
Every edge $v_jv_{j+1}$ belongs to a unique block.
So, think of the partition of the path into maximal segments whose edges belong to the same block.
Whenever consecutive segments meet at a vertex, that vertex belongs to multiple blocks and is therefore a cut vertex.
Since $v_1$ and $v_s$ are cut vertices as well, every segment lies in a block that contains two cut vertices.
By Lemma~\ref{lem:bull-free_structure:two_cutvertices}, every such block is an edge.
Hence, every segment consists of a single edge from $\cD$, and the complete $c,c'$-path lies in $h$.
Thus, $h$ is connected.

Moreover, $h$ cannot contain a cycle.
Indeed, every cycle is a $2$-connected subgraph and must therefore be part of a single block of $G$.
All blocks represented by edges of $h$ belong to $\cD$ and are single edges.
Consequently, $h$ is a tree.

We construct the tree $H$ from $h$ by adding, for every $i \in \{1,\dots,t\}$, a new leaf $x_i$ adjacent to $c_i$.
Suppose that $H$ contained a leaf that is not one of the added vertices $x_i$, that is, some vertex $c \in V(h)=V(H)-\{x_1,\dots,x_t\}$ were a leaf of $H$.
Then $c$ would also be a leaf of $h$.
Since $c$ is a cut vertex of $G$, it belongs to at least two blocks of $G$.
As $c$ has degree one in $h$, only one of these blocks is a duo-block from $\cD$.
Therefore, $c$ must also be the unique cut vertex of some block in $\cS$.
Hence, $c=c_i$ for some $i$, and the additional neighbor $x_i$ gives $c$ degree at least two in $H$, a contradiction.
Therefore, $L = \{x_1,\dots,x_t\}$ is exactly the leaf set of $H$.
Furthermore, every pair of distinct leaves $x_i,x_j \in L$ has different neighbors $c_i$ and $c_j$, by construction.
Since $h$ contains at least two vertices and every leaf of $h$ receives an additional neighbor, the tree $H$ has at least four vertices.

For every $x_i \in L$, we finally define $G_{x_i}=G'_{c_i}$, which has already been shown a bull-free chordal graph without true twins.
We claim that $G$ is obtained from $H$ by substituting every leaf $x_i \in L$ with $G_{x_i}$.
Indeed, the vertex sets $V(h),V(G_{x_1}),\dots,V(G_{x_t})$ form a partition of $V(G)$.
Each of the corresponding induced subgraphs therefore already contains all edges of $G$ whose endpoints belong to the same part.
There are no edges between $G_{x_i}$ and $G_{x_j}$ for distinct $i,j$.
Moreover, for every $i \in \{1,\dots,t\}$, the only edges between $G_{x_i}$ and the remainder of $G$ are the edges $c_iv$ for $v\in V(G_{x_i})$.
According to Lemma~\ref{lem:bull-free_structure:cutvertices}, all of these edges are present.
This is exactly the adjacency created by substituting the leaf $x_i$ of $H$ with $G_{x_i}$.
Thus, $G$ satisfies Condition~\ref{thm:bull_free_decomposition:Three}.
\qed\end{proof}
See the proof for the other direction of Theorem~\ref{thm:bull_free_decomposition} in the appendix.
This completes our characterization of bull-free {\clean} graphs and we are ready to specialize our findings for the two considered subclasses of these graphs.


\subsection{(Bull, Gem)-Free Conchort Graphs Generalize Chordal Cographs}
\label{sec:bull_gem-free-structure}

The structure of bull-free {\clean} graphs $G$, as described in Theorem~\ref{thm:bull_free_decomposition}, is already restrictive and, yet, it suggests a recursive approach to leaf root construction:
(i) decompose $G$ into the tree $H$ and the graphs $G_x$ at the leaves $x$ of $H$, (ii) recursively create an optimal leaf root $T_x$ for every graph $G_x$, and (iii) somehow combine those leaf roots and $H$ into one optimal leaf root for $G$.
And while this may be the right approach for bull-free {\clean} graphs, the whole recursion tree collapses to at most one level for $(\text{bull},\text{gem})$-free {\clean} graphs.

In the $(\text{bull},\text{gem})$-free setting, either the entire graph $G$ is a chordal cograph or all substituted graphs $G_x$ are.
This happens because cut vertices $c$ are universal in the contained blocks $B$.
Hence, if there was an induced $P_4$ in $B$ then, together with the universal $c$, this would induce a gem in $B$ and, thus, in $G$ -- impossible.
The following lemma formally states this structural condition.
\begin{restatable}{lemma}{bullgemfreedecomposition}\label{lem:bull_gem-free_structure}
  A graph $G$ is a $(\text{bull},\text{gem})$-free {\clean} graph if and only if one of the following is true:
  \begin{enumerate}[label=(\roman*)]
    \item\label{lem:bull_gem-free_structure:One} $G$ is a {\clean} cograph.
    \item\label{lem:bull_gem-free_structure:Two} There is a tree $H$ on at least four vertices where distinct vertices in the leaf set $L$ have different neighbors in $H$ and $G$ results from $H$ by substituting every leaf $x \in L$ with a chordal cograph $G_x$ without true twins.
  \end{enumerate}
\end{restatable}
For a leaf root construction approach, this means that, without further recursion, the algorithm of Le and Rosenke~\cite{LeR24} provides an optimal leaf root for $G$ itself or at least all the graphs $G_x$, after some slight extension.
The details on how to combine these leaf roots into one follow in Section~\ref{sec:leaf-roots}.


\subsection{(Bull, Dart)-Free Conchort Graphs Are Trivial Leaf Powers}
\label{sec:bull_dart-free-structure}

While the additional exclusion of gem makes the structure of bull-free {\clean} graphs manageable for leaf root construction, we subsequently see that forbidding dart collapses their structure to two particularly restricted cases.
\begin{restatable}{lemma}{bulldartfreedecomposition}\label{lem:bull_dart-free_structure}
A graph $G$ is a $(\text{bull},\text{dart})$-free {\clean} graph if and only if one of the following is true:
  \begin{enumerate}[label=(\roman*)]
    \item\label{lem:bull_dart-free_structure:One} $G$ is a connected, bull-free proper interval graph without true twins.
    \item\label{lem:bull_dart-free_structure:Two} $G$ is a tree on at least four vertices.
  \end{enumerate}
\end{restatable}
Leaf root construction for the tree case can be considered solved because it corresponds to the $3$-leaf powers, which have been settled in~\cite{BrandstadtL06}.
There, linear-time recognition and $3$-leaf root construction have been well established.

As Dupr\'e la Tour, Lafond and Ndiaye~\cite{DupreLN26} point out, the complexity of computing the leaf rank of proper interval graphs is unknown.
However, the additional constraints of forbidden induced bulls and true twins lead to the following structural insight.
\begin{restatable}{lemma}{bullfreepropinterval}\label{thm:bull-free_prp_int_path_power}
  If $G$ is a connected bull-free proper interval graph without true twins on $n = |V(G)|$ vertices then $G$ is isomorphic to $P^{\omega(G)-1}_n$.
\end{restatable}
With this result, we have basically finished the $(\text{bull},\text{dart})$-free case of Theorem~\ref{thm:main}.
Indeed, after a linear time true twin reduction of the input, every connected component $G_i$ of the result graph $G$ can be classified among the two cases of Lemma~\ref{lem:bull_dart-free_structure} in linear time by first testing whether $G_i$ is a tree and, otherwise, computing a proper interval ordering with a linear-time recognition algorithm~\cite{CorneilKNOS95}.
Based on this, a $\kappa$-leaf root $T$ of $G$ is feasible for the maximum leaf rank $\kappa$ of the component graphs $G_i$.

If we ignore singletons, the leaf rank of any tree component $G_i$ is three and a compressed $k$-leaf root of $G_i$ can be computed for every $k \geq 3$ in linear time, just as in Step~7 of the algorithm in Section~\ref{sec:leaf-roots}.
A non-trivial proper interval component $G_i$ with $n_i = |V(G_i)|$, which is isomorphic to $P_{n_i}^{d}$ for $d = \omega(G_i)-1$, has a $(d+2)$-leaf root that results from taking a path on $n_i$ vertices as a spine and attaching the vertex $v_j$ of $G_i$ as a leaf to the $j$th spine vertex.
If $n_i < 2d+1$ then $v_d$ and $v_{d+1}$ are universal in $G_i$ and therefore true twins.
Thus, $G_i$ contains $P_{2d+1}^{d}$ as an induced subgraph, which has leaf rank $d+2$ as shown by Brandst\"adt and Wagner~\cite{WagnerB09}, implying that the caterpillar is optimal for $G_i$.
It is straightforward to verify that a compressed $(d+3)$-leaf root is obtained from the caterpillar $(d+2)$-leaf root by increasing the weight of every $(d+1)$st spine edge from one to two, leaving all other edge weights unchanged.
With the extension method $\eta$ from~\cite{LeR24}, we can, thus, get a compressed $k$-leaf root of $G_i$ for every $k \geq d+2$ in linear time.

Thus, setting $\kappa$ to the maximum leaf rank of the components, we can construct compressed $\kappa$-leaf roots $T_i$ for all non-trivial components $G_i$.
These roots can be combined in linear time by joining arbitrary internal nodes of consecutive roots through weighted edges of weight $\kappa+1$.
Singleton components can be added through sufficiently long pendant edges, without changing the parameter.
The resulting tree is a compressed optimal $\kappa$-leaf root of $G$.
Every removed true twin $y$ of a vertex $x$ can be reinserted into $T$ as leaf pendent to the neighbor of $x$ in linear time, possibly after a prior subdivision of a compressed edge incident with $x$.
Thus, an optimal leaf root of every $(\text{bull},\text{dart})$-free chordal graph on $n$ vertices and $m$ edges can be constructed in $\cO(n+m)$ time.

\section{Linear Time Optimal Leaf Root Construction}
\label{sec:leaf-roots}

The aim of the section is to prove Theorem~\ref{thm:main}.
Linear-time optimal leaf root construction for $(\text{bull},\text{dart})$-free and $(\text{dart},\text{gem})$-free chordal graphs has been solved in previous sections.
Only the $(\text{bull},\text{gem})$-free case remains.

We first use Lemma~\ref{lem:bull_gem-free_structure} to develop efficient leaf root construction for $(\text{bull},\text{gem})$-free {\clean} graphs.
Here, Condition~\ref{lem:bull_gem-free_structure:One} is a solved case, again.
Recognition of chordal cographs is in linear time~\cite{Chu08} and optimal leaf root construction~\cite{LeR24}, too.

Consequently, we only need to deal with Condition~\ref{lem:bull_gem-free_structure:Two}, where the input graph $G$ is the result of substituting chordal cographs $G_x$ without true twins into the leaves $x$ of a tree $H$ on at least four vertices.
In the following algorithm, we will assume that the input graph $G$ has been decomposed into the graphs $G_x$ and the tree $H$.
Notice that this is not a limiting restriction towards the proof of Theorem~\ref{thm:main}.
In fact, the decomposition can easily be derived from the block-cut-tree $\text{BCT}$ of $G$, which can be computed in linear time~\cite{Tarjan72}.
Every graph $G_x$ can simply be obtained from the union of those blocks in $\text{BCT}$ that are leaves and incident to the same cut vertex.
In fact, the proof of Theorem~\ref{thm:bull_free_decomposition} describes this already by defining $G_x$ as
$
G'_c= G\left[\bigcup_{B\in\cS_c}\bigl(V(B)-\{c\}\bigr)\right]
$
for the cut vertex $c$ adjacent to $x$.
The remaining blocks in $\text{BCT}$ form the graph $h$, which is introduced in the proof of Theorem~\ref{thm:bull_free_decomposition} and, as stated there, $H$ is obtained from $h$ by inserting the leaf set $L$.
To verify that all $G_x$ are chordal cographs and that $H$ is a tree can be done in linear time.

We are ready to present the optimal leaf root construction algorithm.
The intuitive but simplified idea behind the construction is, for a given parity $p$, to create optimal $k$-leaf roots $T_x$ of parity $p$ for slightly extended versions of the chordal cographs $G_x$ and $T_h$ for the tree $h$ and to glue them together into one optimal $k$-leaf root of parity $p$ for $G$.
The actual optimal leaf root of $G$ is then obtained by taking the better one of both parities.
In the following, we describe the algorithm for given $G$ and parity $p$ in formal detail:

\noindent
Let $p \in \{0,1\}$ be a given parity and let $G$ be a $(\text{bull},\text{gem})$-free {\clean} graph given as a tree $H$ on at least four vertices where distinct leaves have different neighbors and, for every leaf $x$ of $H$, a chordal cograph $G_x$ without true twins.
\begin{enumerate}
  \item For every leaf $x$ of $H$ with neighbor $y$, let $z$ be a new vertex and create $G^+_x$ from $G_x$ by making $y$ universal in $V(G_x)$ and adding the edge $yz$.
  \item For every leaf $x$ of $H$, compute a compressed, parity-optimal leaf root $(T'_x, k_x)=\rho(G^+_x,p)$ with the construction method $\rho$ from~\cite{LeR24}.
  \item Compute $k=\max\{k_x \mid x\text{ is a leaf of }H\}$.
  \item For every leaf $x$ of $H$, compute a compressed $k$-leaf root $T''_x=\eta\left(T'_x,\frac{k-k_x}{2}\right)$ of $G^+_x$ with the extension method $\eta$ in~\cite{LeR24}.
  \item For every leaf $x$ of $H$ with neighbor $y$, compute a compressed $k$-leaf root $T_x$ for the graph $G_x+y$ resulting from $G^+_x$ by removing $z$.
  Therefore, $T_x$ is obtained by taking the smallest subtree of $T''_x$ whose leaf set is $V(G_x) + y$.
  \item Let $L$ be the set of leaves of $H$ and let $h = H-L$.
  \item Construct the compressed $k$-leaf root $T_h$ of $h$:
  \begin{enumerate}
    \item For every vertex $v \in V(h)$, create a new vertex $v'$ in $T_h$ and attach the vertex $v$ as a leaf to $v'$.
    \item For every edge $vw \in E(h)$, connect $v'$ and $w'$ in $T_h$ by an edge with weight $k-2$.
  \end{enumerate}
  \item Construct the compressed $k$-leaf root $T$ of $G$ by combining the trees $T_x$ for all leaves $x \in L$ and $T_h$ as follows:
    
    For every leaf $x$ of $H$ with neighbor $y$, attach $T_x$ to $T_h$ by identifying the edge $vy$ between leaf $y$ and its only neighbor $v$ in $T_x$ with the edge $y'y$ in $T_h$ by identifying the vertices $v$ and $y'$ and, simultaneously, the two copies of $y$, so that the pendant edges $vy$ and $y'y$ coincide.
    In the compressed representation of tree $T_x$, if the pendant edge $vy$ containing $y$ is represented by a weighted edge, this requires to first split off the last edge of weight one.
  \item Return $T$ and $k$.
\end{enumerate}

To relate our approach to that of Le and Rosenke~\cite{LeR24}, we extend their $\rho$-notation and, here, use $(T,k)=\rho(G,p)$ to denote the output of the entire above algorithm, including the decomposition of the input graph $G$ into the graphs $G_x$ and the tree $H$.
The following lemma states the correctness of this algorithm.
\begin{restatable}{lemma}{leafrootconstruction}\label{lem:bull_gem-free_structure:leaf_root}
  Let $G$ be a $1$-separable, $(\text{bull},\text{gem})$-free, {\clean} graph that is not a cograph and let $p\in\{0,1\}$.
  Then $(T, k) = \rho(G, p)$ provides a compressed $k$-leaf root $T$ of $G$ such that $\pi(k)=p$ and $k \leq k'$ for every $k'$-leaf root of $G$ with $\pi(k') = p$.
  The computation takes $\cO(n+m)$ time, where $n=|V(G)|$ and $m=|E(G)|$.
\end{restatable}
The proof of this lemma is straightforward.
First, it needs checking the $T$-distance for all pairs of distinct leaves $x,y \in V(G)$.
Pairs that belong to the same graphs $G_x+y$ or $h$ are handled by the corresponding local leaf roots $T_x$ or $T_h$, while all remaining pairs have distance $>k$ through the subdivided copy of $h$.
Then the optimality of the value $k$ with respect to parity $p$ is established through the argument that every graph $G_x^+$ is isomorphic to an induced subgraph of $G$.
Consequently, every $k'$-leaf root of $G$ with $\pi(k')=p$ also yields a $k'$-leaf root of every $G_x^+$, and therefore $k'$ is at least the locally optimal value computed for each of these graphs.
As $\rho$ chooses $k$ as the maximum of these local values, we obtain $k\leq k'$.
Finally, all local leaf roots $T_x$ are constructed and extended in linear time by the algorithms from~\cite{LeR24}, and the subdivision and gluing along $T_h$ can be carried out in $\cO(n+m)$ time.

Together with the previously settled $(\text{bull},\text{dart})$-free and $(\text{dart},\text{gem})$-free cases, Lemma~\ref{lem:bull_gem-free_structure:leaf_root} completes the proof of Theorem~\ref{thm:main}.
Indeed, after reducing true twins, we consider the connected components $G_i$ of the resulting graph $G$ independently.
For every non-trivial $(\text{bull},\text{gem})$-free {\clean} component $G_i$ and each parity $p\in\{0,1\}$, we can compute a compressed $k_{i,p}$-leaf root with minimum parameter $k_{i,p}$ of parity $p$ in linear time, using the algorithm of~\cite{LeR24} if $G_i$ is a chordal cograph, or the algorithm of this section otherwise.
Let $k_p=\max_i k_{i,p}$ and $\kappa=\min\{k_0,k_1\}$, where the maximum ranges over all non-trivial components.
Choosing a parity $p$ with $k_p=\kappa$, the extension method $\eta$ from~\cite{LeR24} allows us to increase the parameters of all corresponding component roots to $\kappa$ in linear time.
As in Section~\ref{sec:bull_dart-free-structure}, these roots can then be combined by joining arbitrary internal nodes of consecutive roots through weighted edges of weight $\kappa+1$.
Singleton components are added through pendant edges of weight $\kappa+1$, yielding a compressed $\kappa$-leaf root $T$ of $G$.
If all components are singletons, the original graph is a disjoint union of cliques, for which OLR is trivial.

Conversely, every $k$-leaf root of $G$ also induces a $k$-leaf root of each component $G_i$, implying $k\geq k_{\pi(k)}\geq\kappa$.
Thus, $\kappa$ is the leaf rank of $G$ and $T$ is optimal.
Since all reductions, decompositions, local constructions, combinations, and the final reinsertion of true twins can be performed in $\cO(n+m)$ time, an optimal leaf root can be constructed in linear time for all three graph classes stated in Theorem~\ref{thm:main}.

\section{Conclusion}
\label{sec:conclusion}

This paper introduces a linear-time solution to OLR for a substantial generalization of chordal cographs.
The unrestricted OLR problem is $\NP$-hard~\cite{DupreLN26} and, since leaf powers are strongly chordal~\cite{RosenkeLB21}, hardness remains even when restricting attention to this chordal subclass.
A meaningful direction for future research is to investigate the precise borderline between efficiently solvable and intractable cases of this problem.
However, the gap between the classes considered here and general leaf powers is still substantial.
From the standpoint established here, natural next milestones are efficient OLR algorithms for the chordal graph classes obtained by forbidding only one of the bull, dart, and gem, with the gem-free chordal graphs also known as the Ptolemaic graphs.
Another non-trivial research objective, also motivated by the results of this paper, is to determine the complexity of OLR on proper interval graphs or even general interval graphs.

\begin{credits}

  \subsubsection*{\ackname}
The authors thank the reviewers for their helpful comments.
\mbox{OpenAI's} GPT-5.6 Sol assisted in the paper preparation through proof checking, argument polishing, language editing, and bibliographic research.
The authors independently created all mathematical ideas and results except the AI-originated Lemma~\ref{thm:bull-free_prp_int_path_power}, critically verified all AI-assisted material, and take full responsibility for the paper.

\subsubsection*{\discintname}
The authors declare that they have no competing interests.
\end{credits}

\bibliographystyle{splncs04}
\bibliography{leafpowers}

\appendix
\renewcommand*{\theHsection}{appendix.\arabic{section}}
\section{Appendix}
\label{sec:appendix}

\cutvertexblock*
\begin{proof}
  Let $B$ be any block of $G$ that contains $c$.
  If $B$ is simply an edge $cx$ with $x \in V(G)-c$ then $c$ is adjacent to the only other vertex $x$ of the block, and we are done.
  Otherwise, $B$ is a $2$-connected induced subgraph of $G$ with at least two additional vertices aside from $c$.
  Since $c$ is a cut vertex there exists a vertex $x \in N_G(c) - V(B)$.
  For all $v \in V(B)-c$, the vertices $v$ and $x$ are not adjacent because any such edge $vx$ would imply two neighbors of $x$ in $B$, namely $c$ and $v$, which makes $B+x$ a $2$-connected subgraph of $G$, a contradiction to the maximality of $B$.

  For a contradiction, assume that there is at least one vertex in $V(B)-c$ that is not adjacent to $c$.
  In that case, select a vertex $u \in V(B)-N_B[c]$ that minimizes the distance $d_B(u,c)$ and let $u = v_1, \dots, v_s = c$ be a shortest $u,c$-path.
  Being shortest, the path is also induced and, hence, $s = 3$, as, otherwise, $v_2 \in V(B)-N_G[c]$ and $d_G(v_2,c) < d_G(u,c)$.
  For brevity, we let $v = v_2$ and get an induced path $u, v, c, x$ in $G$.

  However, since $B$ is $2$-connected, $v$ is not a cut vertex of $B$ and, so, there is at least one other $u,c$-path within $B-v$ that omits $v$.
  Among those paths, we let $u = w_1, \dots, w_t = c$ be a shortest one.
  Together with $v$, this path creates a cycle $v, w_1, \dots, w_t, v$ in $G$ that, as $G$ is chordal, has at least one chord.
  However, chords $w_iw_j$ for $1 \leq i<j \leq t$ cannot exist because they would create a shorter $u,c$-path $u = w_1, \dots, w_i, w_j, \dots, w_t = c$ that omits $v$.
  Hence, every chord of the cycle must be incident to $v$.
  In fact, $v$ has to be adjacent to all of $w_1, \dots, w_t$.
  If $s'$ were the smallest integer in $\{2, \dots, t-1\}$ such that $vw_{s'} \notin E(G)$ then let $t' \in \{s'+1, \dots, t\}$ be the smallest integer such that $vw_{t'} \in E(G)$ and observe the induced cycle $v, w_{s'-1}, w_{s'}, \dots, w_{t'-1}, w_{t'}, v$ of length $t'-s' + 3$.
  Since $s' < t'$ this induced cycle cannot exist in the chordal graph $G$ and, hence, $vw_1, \dots, vw_t \in E(G)$.
  Based on this, we see that $t = 3$ as, otherwise, the induced $P_4$ on $u, v, c, x$ and the tip vertex $w_{t-1}$ would induce a bull in $G$.
  Again, we let $w = w_2$, for short, and obtain an induced $P_4$ on $u, v, c, x$ and an additional vertex $w$ that sees $u,v,c$ but not $x$.

  Next recall that $v$ and $w$ cannot be true twins and, since $u,c,x$ do not distinguish them, there has to be a vertex $y \in V(G)-\{u,v,w,c,x\}$ that, without loss of generality, is adjacent to $v$ but not to $w$.
  If $y$ was neither adjacent to $c$ nor $x$ then there would be an induced bull in $G$ with $P_4$ on $y,v,c,x$ and tip $w$.
  This is impossible and so is $y$ being adjacent to $x$ but not to $c$ as this would induce the cycle $y,v,c,x,y$, a $C_4$.
  Therefore, $cy \in E(G)$ and, since every edge belongs to a unique block, we get that the block containing the triangle $c,v,y$ must be the block $B$, which hosts the edge $cv$.
  Hence, $y \in V(B)$ and, consequently, $xy \notin E(G)$.
  But this means that the presence of edge $uy$ creates an induced cycle $y,c,w,u,y$ of length four and its absence creates an induced bull with $P_4$ on $u,v,c,x$ and tip vertex $y$, both of which are impossible.
  This contradiction to the existence of $u$ completes the proof.
\qed\end{proof}


\twocutvertices*
\begin{proof}
  Let $B$ be a block of $G$.
  If $B$ is an edge then there is nothing to prove.
  Otherwise, assume for a contradiction that $B$ contains at least two distinct cut vertices $u$ and $v$ and at least one more vertex $w$.
  According to Lemma~\ref{lem:bull-free_structure:cutvertices}, the vertices $u,v,w$ form a triangle.
  
  Since $u$ and $v$ are cut vertices there exist vertices $x \in N_G(u) - V(B)$ and $y \in N_G(v) - V(B)$.
  Like in the proof of Lemma~\ref{lem:bull-free_structure:cutvertices}, we notice for every vertex $z \in V(B)-u$ that $x$ and $z$ are not adjacent and, equivalently, none of the vertices $z \in V(B)-v$ is adjacent to $y$.
  As $ux, vy \in E(G)$, this immediately implies $x \neq y$.
  Therefore, $u,v,w$ is a triangle in $G$ and $x,y$ are two vertices with $u = N_G(x) \cap \{u,v,w\}$ and $v = N_G(y) \cap \{u,v,w\}$.

  Now, the presence of the edge $xy$ creates an induced cycle $x,u,v,y,x$ of length four and the absence of that edge creates an induced bull with $P_4$ on $x,u,v,y$ and tip vertex $w$.
  Both lead to a contradiction, which completes the proof.
\qed\end{proof}


\bullfreedecomposition*
\begin{proof}
The $\Rightarrow$-direction has already been shown above.

For the $\Leftarrow$-direction, suppose first that $G$ satisfies Condition~\ref{thm:bull_free_decomposition:One}.
Then $G$ is a bull-free {\clean} graph and we are done.

\medskip
Otherwise, if $G$ satisfies Condition~\ref{thm:bull_free_decomposition:Two}, let $u$ be its universal cut vertex, and let $G_1,\dots,G_t$ be the connected components of $G-u$, which are bull-free {\clean} graphs.
Since $G-u$ is the disjoint union of the graphs $G_i$, the graph $G-u$ is bull-free and chordal.
Adding the universal vertex $u$ can neither create an induced cycle of length at least four nor an induced bull and, hence, $G$ is bull-free and chordal, too.

Two vertices in distinct components of $G-u$ are non-adjacent and therefore cannot be true twins.
Since every $G_i$ is true twin-free, adjacent vertices $x$ and $y$ in the same component $G_i$ are distinguished by a third vertex in $G_i$ and, hence, $x$ and $y$ are not true twins in $G$.
Furthermore, since $u$ is a cut vertex, we have $t \geq 2$.
Thus, for every $x \in V(G_i)$, there is a vertex in some component $G_j$, where $j\neq i$, that is adjacent to $u$ but not to $x$.
Hence, $u$ and $x$ are not true twins, either.
Consequently, $G$ is a bull-free {\clean} graph.

\medskip
Finally, suppose that Condition~\ref{thm:bull_free_decomposition:Three} holds and let $G$ be the result of substituting bull-free chordal graphs $G_{x_1}, \dots, G_{x_t}$ without true twins into the leaf set $L = \{x_1, \dots, x_t\}$ of a suitable tree $H$ on at least four vertices.
For every $i \in \{1,\dots,t\}$, let $c_i$ denote the unique neighbor of $x_i$ in $H$.
Since $H$ is connected, every graph $G_{x_i}$ is non-empty, and the substitution creates all edges between $c_i$ and $V(G_{x_i})$, the graph $G$ is connected.

The tree $H$ is bull-free and chordal.
Moreover, every edge between two vertices of $H-L$ remains a bridge in $G$, and every graph $G_{x_i}$ is attached to the remainder of $G$ only through $c_i$.
Consequently, every $2$-connected subgraph of $G$ with at least three vertices is contained in $G_{x_i}+c_i$ for some $i \in \{1,\dots,t\}$.

Suppose that $G$ contains an induced cycle of length at least four.
As such a cycle is $2$-connected, it is contained in $G_{x_i}+c_i$ for some $i$.
Since $c_i$ is universal in $V(G_{x_i})$, it cannot belong to an induced cycle of length at least four.
Thus, the complete cycle is contained in $G_{x_i}$, a contradiction to the chordality of $G_{x_i}$.
Hence, $G$ is chordal.

Now suppose that $G$ contains an induced bull with $P_4$ on $a,b,d,e$ and tip vertex $c$.
The triangle on $b,c,d$ is $2$-connected and, by the same argument, is contained in $G_{x_i}+c_i$ for some $i$.
The vertex $c_i$ cannot be one of $b,c,d$, because every vertex of $G_{x_i}$ is adjacent to $c_i$, while vertices outside $G_{x_i}$ have no neighbors in $G_{x_i}$ except through $c_i$, which makes the required adjacencies of the pendant vertices $a$ and $e$ impossible.
Moreover, $c_i$ cannot be $a$ or $e$, since it is adjacent to all three vertices of the triangle.
Thus, $c_i$ is none of the five vertices of the bull.
It follows that all five vertices belong to $G_{x_i}$ and induce a bull there, a contradiction.
Therefore, $G$ is bull-free.

It remains to show that $G$ is true twin-free.
Suppose, to the contrary, that $G$ contains true twins $y$ and $z$.
Clearly, $y,z \in V(G_{x_i})$ for the same $x_i \in L$ would mean that they are true twins in $G_{x_i}$, which is impossible by assumption.
Next, let $y \in V(G_{x_i})$ and $z \in V(G_{x_j})$ for distinct $i,j \in \{1,\dots,t\}$.
Since true twins are adjacent, we would have $yz \in E(G)$, which is impossible, too.

We can also exclude that both $y$ and $z$ belong to $V(H)-L$.
Indeed, true twins are adjacent, and two adjacent vertices of a tree can have the same closed neighborhood only if the tree consists of the single edge $yz$.
Any additional neighbor in $H$, or any vertex substituted for an additional leaf, distinguishes the two vertices in $G$.
This contradicts the fact that $H$ has at least four vertices.

So consider the last possibility where, without loss of generality, $y \in V(G_{x_i})$ and $z \in V(H)-L$.
Since $y$ and $z$ are adjacent, we have $z=c_i$.
As $H$ has at least four vertices, the vertex $c_i$ has a neighbor in $H$ other than $x_i$.
By the assumption that distinct leaves of $H$ have different neighbors, such a neighbor belongs to $V(H)-L$.
It is adjacent to $z=c_i$ but not to $y$, contradicting that $y$ and $z$ are true twins.

Therefore, $G$ is true twin-free and, thus, a bull-free {\clean} graph, which completes the proof.
\qed\end{proof}


\bullgemfreedecomposition*

\begin{proof}
We begin with the $\Rightarrow$-direction and, therefore, let $G$ be a $(\text{bull},\text{gem})$-free {\clean} graph.
Since $G$ is connected and true twin-free, it is either a single vertex, $2$-connected or $1$-separable.
We argue that, in the first two cases, $G$ is a chordal cograph, which is trivial for a single vertex.

Assume for a contradiction that $G$ has an induced $P_4$, say $a,b,c,d$.
Since $G$ is $2$-connected and chordal, every edge, including $bc$, is part of a triangle.
This happened because $b$ and $c$ are, beside the direct edge $bc$, connected by at least one alternate path that avoids the edge $bc$.
If, among them, we select the shortest path then chordality makes sure that the path consists of just two edges, which form a triangle with $bc$.
Say that there is a triangle $b,x,c$.
The vertex $x$ cannot be the tip of an induced bull and, so, at least one of the edges $ax$ and $dx$ exists.
That these edges cannot be present simultaneously follows from the fact that, otherwise, the same $P_4$ and the same tip induce a gem.
By symmetry, we may assume without loss of generality that $ax \in E(G)$ and $dx \notin E(G)$.
  
As $G$ is free of true twins, there is a vertex $y$ that distinguishes the neighborhoods of $b$ and $x$.
Again, symmetry allows us to assume without loss of generality that $y \in N_G(b) - N_G[x]$.
If none of the edges $ay, cy, dy$ is present in $G$, there would be an induced bull with $P_4$ on $y,b,c,d$ and tip $x$.
That $ay \notin E(G)$ is easily checked since, otherwise, $cy \notin E(G)$ would imply an induced gem with $P_4$ on $y,a,x,c$ and tip $b$ and with $cy \in E(G)$ there would be an induced $C_4$ on $y,a,x,c,y$.
Similarly, $cy \in E(G)$ would produce an induced gem with $P_4$ on $a,x,c,y$ and tip $b$.
This means that just $dy \in E(G)$.
However, this induces a $C_4$ on $y,b,c,d,y$, which delivers the final contradiction to the initial assumption.

This means that $G$ is a chordal cograph and, thus, satisfies Condition~\ref{lem:bull_gem-free_structure:One}.
Recall, however, that every connected chordal cograph without true twins is either a single vertex or has a universal cut vertex.
Since the single vertex case has been excluded before, the $2$-connected case handled above cannot really occur.

\smallskip
Now, let $G$ be a $1$-separable graph.
According to Theorem~\ref{thm:bull_free_decomposition}, $G$ either has a unique universal cut vertex or is the result of substituting every leaf $x$ of a suitable tree $H$ on at least four vertices with a bull-free, true twin-free graph $G_x$.

First, if $G$ has a universal cut vertex $u$ then every connected component of $G-u$ is $P_4$-free.
If there was an induced $P_4$ in one of these components then, with the universal vertex $u$, there would also be an induced gem in $G$, a contradiction.
Since $G-u$ is an induced subgraph of $G$ this means that $G-u$ is a chordal cograph.
The full join between $G-u$ and the vertex $u$ preserves both properties.
This implies that $G$ is a chordal cograph and, thus, Condition~\ref{lem:bull_gem-free_structure:One}.
  
Second, if $G$ is the result of the graph substitutions then suppose there is a leaf $x \in V(H)$ for which $G_x$ is not a chordal cograph, thus, contains an induced $P_4$, say on $a,b,c,d$.
In the tree $H$, let $y$ be the neighbor of leaf $x$.
Due to the substitution of $x$ with $G_x$ in $H$, we have $vy \in E(G)$ for all $v \in V(G_x)$.
This includes the vertices $a,b,c,d$ and, thus, $\{a,b,y,c,d\}$ induces a gem in $G$, a contradiction.
Therefore, every $G_x$ is a chordal cograph and Condition~\ref{lem:bull_gem-free_structure:Two} holds.
This completes the first direction.
  
\medskip
For the $\Leftarrow$-direction, we start with Condition~\ref{lem:bull_gem-free_structure:One}, where $G$ is a {\clean} cograph.
By definition, this means that $G$ is $P_4$-free, which, in particular, excludes induced bulls and gems.
  
Under Condition~\ref{lem:bull_gem-free_structure:Two}, $G$ results from a suitable tree $H$ by substituting the leaves $x$ with chordal cographs $G_x$ without true twins.
Since bull contains an induced $P_4$, the graphs $G_x$ are bull-free and, thus, Theorem~\ref{thm:bull_free_decomposition} implies that $G$ is bull-free {\clean} graph.

It remains to show that $G$ is gem-free.
So assume, for a contradiction, that $G$ has an induced gem, say with $P_4$ on $a,b,d,e$ and tip $c$.
Each cycle of the gem would entirely be contained in one block of $G$ -- in particular, the cycle $c,a,b,d,e,c$ along the gem's $P_4$ and around the tip $c$.
Hence, the entire gem is contained in a block $B$ that is a subgraph of $G_x + y$ for some $H$-leaf $x$, where $y$ is the neighbor of $x$ in $H$.
However, $G_x$ is a chordal cograph and $y$ is complete for $V(G_x)$, which makes the graph $G_x+y$ and, thus, the block $B$ a chordal cograph.
However, $B$ contains the induced $P_4$ on $a,b,d,e$, which is a contradiction.
Therefore, $G$ is $(\text{bull},\text{gem})$-free {\clean} graph, which completes the proof.
\qed\end{proof}


\bulldartfreedecomposition*
The following proof references a number of small graphs that are not relevant before.
Figure~\ref{fig:more_small_graphs} introduces them.
\begin{figure}[!ht]
  \centerline{
  \begin{tikzpicture}[scale = \myscale, every node/.style={scale=\myscale}]
	\begin{scope}[xshift=0cm]
    	\drawGraphlet{a/{0,0}/above,b/{1.5,0}/above,c/{3,0}/above,d/{0.75,-0.66}/below left,e/{2.25,-0.66}/below right,f/{1.5,-1.33}/below left}{a/b,b/c,a/d,b/d,b/e,c/e,d/e,d/f,e/f}{1}
      \node at (1.5,1.125) {(i)};
	\end{scope}
	\begin{scope}[xshift=5cm]
    	\drawGraphlet{a/{0,0}/above,b/{1,0}/above,e/{1.5,-0.66}/below left,c/{2,0}/above,d/{3,0}/above,f/{1.5,-1.33}/below left}{a/b,b/c,c/d,b/e,c/e,e/f}{1}
      \node at (1.5,1.125) {(ii)};
	\end{scope}
  \begin{scope}[xshift=10cm]
    	\drawGraphlet{a/{0,0}/left,b/{3,0}/right,c/{1.5,-1.33}/below,x/{1.5,-0.66}/above}{a/x,b/x,c/x}{1}
      \node at (1.5,1.125) {(iii)};
	\end{scope}
  \end{tikzpicture}}
  \caption{(i) $3$-sun, (ii) net, (iii) claw}
  \label{fig:more_small_graphs}
\end{figure}
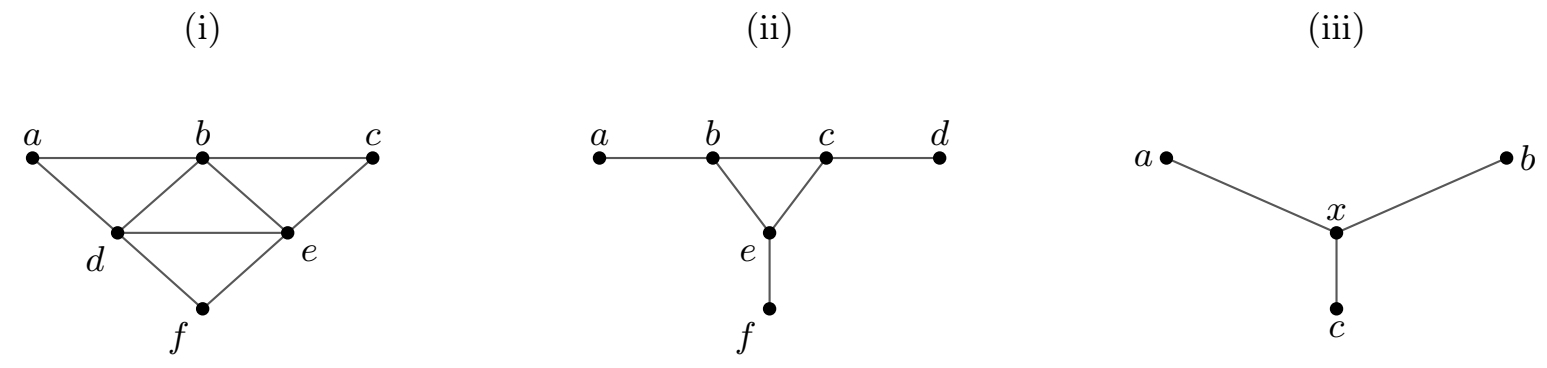
\begin{proof}[of Lemma~\ref{lem:bull_dart-free_structure}]
  As usual, we start with the $\Rightarrow$-direction and let $G$ be a $(\text{bull},\text{dart})$-free {\clean} graph.
  That $G$ is connected and true twin-free implies again that it is either a single vertex, $2$-connected or $1$-separable.
  
  In the first two cases, $G$ is a proper interval graph as we see in the following.
  As a chordal graph, $G$ is proper interval, if and only if it is free of $3$-sun, net and claw~\cite{Wegner1967}.
  This is trivially true for graphs that consist of just one vertex.
  Moreover, bull already forbids $3$-sun and net, which both contain induced bull.
  This leaves us with claw.
  We show that forbidding dart in a $2$-connected chordal graph without true twins already eliminates claw.
  
  So, for a contradiction, assume an induced claw in $G$ with center $x$ and leaves $a,b,c$.
  Since $G$ is $2$-connected there are $v,w$-paths for all distinct $v,w \in \{a,b,c\}$ that avoid $x$.
  For any pair, consider a shortest $v,w$-path that avoids $x$ and notice that none of the three paths can be the simple edge $vw$ because the claw is induced.
  Every selected $v,w$-path $v = v_1, \dots, v_t = w$ creates a cycle $x,v_1,\dots,v_t,x$ in $G$.
  In a chordal graph, this cycle has a chord and, in fact, we argue that the chord $v_ix \in E(G)$ is present for all $i \in \{2, \dots, t\}$.
  Indeed, the chord $v_iv_j$ cannot be present for any $i,j \in \{1, \dots, t\}$ with $i+1 < j$ because this would imply a shorter $v,w$-path $v_1,\dots,v_i,v_j,\dots,v_t$.
  Therefore, all chords have to be incident with $x$.
  Now, if $v_ix$ and $v_jx$ are edges for any $1 \leq i < i+1 < j \leq t$ and the chords $v_{i+1}x, \dots, v_{j-1}x$ are missing then this would induce a cycle $x,v_i,\dots,v_j,x$ of length at least four.
  Consequently, $x$ has to be adjacent to all vertices in the path.

  Assume without loss of generality that the length $s-1$ of the $a,b$-path $a = u_1, \dots, u_s = b$ is shortest among the three selected paths.
  Clearly $s \geq 3$ and we show that $s = 3$.
  Indeed, if $s > 3$ then $c$ cannot be adjacent to any vertex $u_i \in \{u_2, \dots, u_{s-1}\}$ because this would imply the existence of an $x$-avoiding $a,c$- or $b,c$-path of length $\min\{i, s-i+1\} < s-1$, a contradiction to the choice of the $a,b$-path.
  But then there is an induced dart with $P_3$ on $u_2,x,c$ and $u_1,u_3$ the wing tips.
  This is a contradiction and, thus, if we rename $y = u_2$ then the path is $a, y , b$.

  That $cy \in E(G)$ is obvious because, otherwise, we still have the same induced dart as before.
  The vertices $x$ and $y$ are not true twins in $G$ and, so, without loss of generality, there is a vertex $z \in N(y) - N[x]$ which implies $z \notin \{a,b,c,x,y\}$.
  Now, for every pair $v,w \in \{a,b,c\}$ it must be true that $z$ is adjacent to exactly one, $v$ or $w$.
  If $z$ sees neither $v$ nor $w$ then there is an induced dart with $P_3$ on $x,y,z$ and $v,w$ the wing tips and if $z$ is adjacent to both then $z,v,x,w,z$ is an induced $C_4$.

  But, on the one hand, $z$ seeing at least one vertex from every pair means that $z$ is adjacent to at least two of $a,b,c$.
  On the other hand, seeing at most one of each pair implies that $z$ is adjacent to at most one of $a,b,c$.
  These statements contradict each other and, hence, the induced claw cannot exist, which makes Condition~\ref{lem:bull_dart-free_structure:One} true.

  \smallskip
  Next, let $G$ be a $1$-separable graph.
  Using Theorem~\ref{thm:bull_free_decomposition}, we know that $G$ has a unique universal cut vertex or results from the substitution of a bull-free, true twin-free graph $G_x$ into every leaf $x$ of a suitable tree $H$ of at least four vertices.

  If $G$ has a universal cut vertex $u$ then consider the connected components $G_1, \dots, G_t$ of $G-u$, where $t \geq 2$.
  Assume, for a contradiction, that any $G_i$ contained an induced $P_3$, say on $a,b,c$.
  Then we select $d \in V(G_j)$ for any $j \neq i$ and observe that, with the central universal vertex $u$, there would be a dart in $G$ induced by $\{a,b,c,u,d\}$.
  Since they are connected, $G_1, \dots, G_t$ are therefore complete graphs.
  If any $G_i$ contains distinct vertices $v$ and $w$ then $N_G[v] = V(G_i)+u = N_G[w]$, which implies the existence of true twins $v, w$ in $G$.
  Since this is impossible, $G_1, \dots, G_t$ are single vertices and $G$ is a star.
  If the star has just two leaves then Condition~\ref{lem:bull_dart-free_structure:One} is satisfied, otherwise Condition~\ref{lem:bull_dart-free_structure:Two}.

  If $G$ results from the graph substitution into the leaves of $H$, the same arguments show that all the graphs $G_x$ are $P_3$-free.
  Since they are free of true twins but do not have to be connected, we see that they are independent sets.
  Hence, here $G$ is a tree that results from $H$ by replacing every leaf $x$ with one or multiple leaves.
  So in this case, Condition~\ref{lem:bull_dart-free_structure:Two} holds.

  \medskip
  Now for the $\Leftarrow$-direction.
  If Condition~\ref{lem:bull_dart-free_structure:One} is true then $G$ is a connected bull-free proper interval graph without true twins.
  Since interval graphs are chordal, this already means that $G$ is bull-free {\clean}.
  But proper interval graphs are claw-free, too, and, since dart has an induced claw, we get that $G$ is a $(\text{bull},\text{dart})$-free {\clean} graph.

  Otherwise, under Condition~\ref{lem:bull_dart-free_structure:Two}, $G$ is a tree.
  Since trees of at least four vertices are connected, acyclic graphs without true twins and because bull and dart contain cycles, it follows that $G$ is $(\text{bull},\text{dart})$-free {\clean}, again.
  This completes the proof.
\qed\end{proof}


\bullfreepropinterval*

Before we give the proof, we introduce some further terminology related to proper interval graphs that is not relevant before this point.
In the following, we fix an ordering $<_G$ for every proper interval graph considered. 
For a vertex $v$, let
\[
  \bot_G(v)=\min_{<_G}N_G[v]
  \qquad\text{and}\qquad
  \top_G(v)=\max_{<_G}N_G[v].
\]
Moreover, $\min_G$ and $\max_G$ denote the first and the last vertex of $G$, respectively, with respect to $<_G$.
If $v$ has a successor, respectively predecessor, in this ordering then it is denoted by $v+1$, respectively $v-1$.
More generally, $v+i$ denotes the vertex obtained by moving $i$ positions to the right of $v$, whenever it exists.
Proper interval orderings satisfy the interval monotonicity
\[
  u\leq_G v
  \quad\Longrightarrow\quad
  \bot_G(u)\leq_G\bot_G(v)
  \text{ and }
  \top_G(u)\leq_G\top_G(v).
\]
With this notation, we can prove the Lemma.
\begin{proof}[of Lemma~\ref{thm:bull-free_prp_int_path_power}]
  Let $G$ be a connected bull-free proper interval graph without true twins.
  The proof works in three steps -- the first one is the following claim.

  \begin{claim}\label{lem:bull-free_prp_int_neighborhood}
    For all distinct vertices $v$ and $w$ in $G$, the relation $N_G[v] \subseteq N_G[w]$ holds only if $N_G[v]$ contains $\min_G$ or $\max_G$.
  \end{claim}
    Suppose, towards a contradiction, that there are distinct vertices $v$ and $w$ in $G$ such that $N_G[v] \subseteq N_G[w]$ but $\{\min_G,\max_G\} \cap N_G[v] = \emptyset$.
    Without true twins in $G$, we then even have $N_G[v] \subset N_G[w]$.
    
    In the first case, we let $v <_G w$, which implies $\bot_G(v) \leq_G \bot_G(w)$ and $\top_G(v) \leq_G \top_G(w)$.
    That $N_G[v] \subset N_G[w]$ also means that $\bot_G(v) \geq_G \bot_G(w)$ and $\top_G(v) <_G \top_G(w)$.
    Hence, $\bot_G(v) = \bot_G(w)$ and, if we let $u = \bot_G(v)$ then $u,v,w$ form a triangle in $G$.
    Since $\min_G \notin N_G[v]$, we can select the immediate predecessor $x$ of $u$ (the largest vertex of $G$ with respect to $<_G$ that fulfills $x <_G u$).
    Since $G$ is connected and $x$ and $u$ are consecutive with respect to $<_G$, we have $xu \in E(G)$.
    If we let $y = \top_G(w)$ then, by construction, the triangle $u,v,w$ has incident edges $xu$ and $wy$.
    Moreover, since $G$ is a proper interval graph and $x <_G u = \bot_G(v) = \bot_G(w) <_G \top_G(v) <_G \top_G(w) = y$, we know that $xv, xw, uy, vy, xy \notin E(G)$.
    This induces a bull in $G$ with $P_4$ on $x,u,w,y$ and tip $v$.

    The second case, $v >_G w$, works by a reduction to the first, if we simply invert the ordering $<_G$.
    In this way, we can use the same but \emph{mirrored} arguments, where $y = \bot_G(w) <_G \bot_G(v)$ and $u = \top_G(w) = \top_G(v)$, and $x$ is the immediate successor of $u \in N_G[v]$, which exists since $\max_G \notin N_G[v]$ and is adjacent to $u$ since consecutive vertices of a connected proper interval graph are adjacent.
    This completes the proof of the claim.

  \medskip
  The second step is formulated in the following claim, which builds on the first.
  \begin{claim}\label{lem:bull-free_prp_int_top_successor}
    For every vertex $v \in V(G) - N_G[\max_G]$, it is true $\top_G(v)+1 = \top_G(v+1)$.
  \end{claim}
    Consider any $v \in V(G) - N_G[\max_G]$ and notice that this implies the existence of a successor $v+1$.
    By interval monotonicity, we have $\bot_G(v) \leq_G \bot_G(v+1)$ and $\top_G(v) \leq_G \top_G(v+1)$.
    
    Assume first that $\top_G(v+1) = \top_G(v)$.
    By precondition, this means $v+1 \notin N_G[\max_G]$.
    Moreover, $\bot_G(v) <_G \bot_G(v+1)$ as, otherwise, $v$ and $v+1$ would be true twins.
    While this implies that $\min_G \notin N_G[v+1]$, it also means that $N_G[v+1] \subseteq N_G[v]$, which is prohibited due to the first claim in this proof. 
    Hence, the assumption fails.

    Secondly, assume for $x = \top_G(v)+1$ and $y = x+1$ that $y \leq_G \top_G(v+1)$.
    Hence, neither $x$ nor $y$ are adjacent to $v$, but both are adjacent to $v+1$.
    Since $v$ and $v+1$ are consecutive and $G$ is a proper interval graph $\bot_G(x) = \bot_G(y) = v+1 \neq \min_G$.
    Again, interval monotonicity and the absence of true twins implies $\top_G(x) <_G \top_G(y)$.
    Like above, this implies both, $\max_G \notin N_G[x]$ and $N_G[x] \subseteq N_G[y]$, which is again impossible according to the first claim in this proof. 
    Therefore, this assumption fails, too, and the only remaining case is $\top_G(v+1) = \top_G(v)+1$.
    This proves the claim.

  \medskip
  The third step proves the lemma and is based on second step.

  Assume that $v_1 <_G \dots <_G v_n$ and let $d$ such that $v_1 + d = \top_G(v_1)$.
  We argue via induction on $i \in \{1, \dots, n-1\}$ that for all $j \in \{i+1, \dots, n\}$ it is true $v_iv_j \in E(G)$ if and only if $j-i \leq d$.
  Clearly, this is true at the induction start $i = 1$, where exactly every $j \in \{2, \dots, d+1\}$ satisfies
  \[
    v_j \leq_G \top_G(v_1) = v_{d+1} \Rightarrow j - i \leq d.
  \]

  Now, for the induction step, let $i > 1$ and assume the statement true for all indices less than $i$.
  If $v_{i-1} \in N_G[\max_G]$ then $\{v_{i-1}, v_i, \dots, v_n\}$ is a clique and $j-i \leq n-i < n-(i-1) \leq d$ for all $j \in \{i+1, \dots, n\}$.
  
  Otherwise, the second claim in this proof 
  states that $\top_G(v_{i-1})+1 = \top_G(v_i)$.
  Since $v_{i-1} \notin N_G[\max_G]$, the induction hypothesis implies $\top_G(v_{i-1}) = v_{i-1} + d$.
  Hence, for all $j \in \{i+1, \dots, n\}$, we get
  \begin{align*}
    v_iv_j \in E(G) &\Leftrightarrow v_j \leq_G \top_G(v_i) \Leftrightarrow v_j \leq_G \top_G(v_{i-1}) + 1\\
    &\Leftrightarrow v_j \leq_G v_{i-1} + d + 1 \Leftrightarrow v_j \leq_G v_i + d \Leftrightarrow j - i \leq d.
  \end{align*}
  In short, this means for all $v_i, v_j$ with $1 \leq i < j \leq n$ that $v_iv_j \in E(G)$ if and only if $j-i \leq d$, or in other words, $G$ is isomorphic to $P_n^{d}$.

  \smallskip
  It remains to show that $d = \omega(G)-1$.
  Consider any clique $K$ of $G$ with maximum cardinality and let $v_i$ and $v_j$ be the minimum, respectively, the maximum in $K$ with respect to $<_G$.
  Then $\top_G(v_i) = v_j$ and hence, $j-i = d$.
  Since $K$ is a maximum clique, there are exactly $\omega(G)-2$ vertices between $v_i$ and $v_j$ and, hence, $d = j-i = \omega(G)-1$. 
\qed\end{proof}


\leafrootconstruction*
\begin{proof}
  Let $H$ be a tree on at least four vertices where distinct leaves have different neighbors and, for every leaf $x$ of $H$, let $G_x$ be chordal cograph without true twins such that $G$ results from $H$ by substituting the graph $G_x$ for every leaf $x$.
  
  Let $x$ be a leaf of $H$ and $y$ be the neighbor of $x$ in $H$.
  We collect some properties of the graph $G^+_x$.
  \begin{claim}
    The graph $G^+_x$ is isomorphic to an induced subgraph of $G$.
  \end{claim}
    Since $H$ has at least four vertices, $y$ has another neighbor aside from $x$, say $q$.
    According to precondition, $q$ is not a leaf of $H$ and, thus, a vertex of $G$.
    Consequently in $G$, the vertex $q$ is adjacent to $y$ and has no neighbor in $V(G_x)$.
    Hence, $G^+_x$ is isomorphic to the induced subgraph
    $
      G_x + \{y,q\} = G[V(G_x) + \{y,q\}]
    $
    of $G$, where $q$ takes the role of the artificial vertex $z$.
    This proves the claim.

  \begin{claim}
    The graph $G^+_x$ is a connected chordal cograph without true twins and universal vertex $y$.
  \end{claim}
    Since $G_x$ is substituted for $x$, we get for all $v \in V(G_x)$ that the edge $yv$ is present in $G$ and, due to construction, in $G^+_x$.
    This is called a \emph{full join} between $y$ and $V(G_x)$, and it makes $y$ universal in $V(G_x)$.
    In the construction of $G^+_x$, we also made $y$ and the artificial vertex $z$ adjacent.
    Thus, the graph $G^+_x$ is connected and $y$ is universal in $G^+_x$.
    
    Moreover,
    $
      G^+_x-y
    $
    is the disjoint union of the chordal cograph $G_x$ and the isolated vertex $z$.
    Since the class of cographs is closed under disjoint union and full join~\cite{CorneilLB81}, $G^+_x$ is a cograph.
    Adding universal or degree-one vertices to a chordal graph preserves chordality, which makes $G^+_x$ chordal, too.
    
    Finally, $G^+_x$ has no true twins.
    Indeed, two vertices in $G_x$ cannot be true twins by assumption, $z$ is adjacent only to $y$, and $z$ distinguishes $y$ from every vertex in $G_x$.
    Thus, $G^+_x$ is a connected chordal cograph without true twins and universal vertex $y$, which settles the claim.

  \medskip
  Le and Rosenke~\cite{LeR24} show that $(T'_x,k_x)=\rho(G^+_x,p)$ is a $k_x$-leaf root of $G^+_x$ with $\pi(k_x)=p$ and minimum $k_x$ among all leaf roots of $G^+_x$ whose parameter has parity $p$.
  By the definition of $k$, we have $k \geq k_x$ for every $x \in L$.
  Since all $k_x$ have parity $p$, the number
  $
    \delta_x=\frac{k-k_x}{2}
  $
  is a non-negative integer (possibly $0$).
  Therefore, 
  $
    T''_x = \eta(T'_x,\delta_x)
  $
  is a $k$-leaf root of $G^+_x$.
  Taking the smallest subtree whose leaf set is $V(G_x) + y$ does not change any distance between these leaves.
  Hence, $T_x$ is a $k$-leaf root of $G_x + y$.

  We need one more observation about $T_x$.
  \begin{claim}
    For every $v \in V(G_x)$, we have $\dist_{T_x}(v,y) \geq 3$.
  \end{claim}
    That $\dist_{T_x}(v,y) \leq 1$ is impossible because $v$ and $y$ are leaves of the tree $T_x$.
    Suppose that $\dist_{T_x}(v,y) = 2$.
    Since $T_x$ is a subtree of $T''_x$, the same equality would hold in that tree, and $v$ and $y$ therefore would have the same neighbor in $T''_x$.
    Consequently,
    $
      \dist_{T''_x}(v, w)=\dist_{T''_x}(y, w)
    $
    for every vertex $w \in V(G^+_x) - \{v,y\}$.
    Moreover, $vy$ would be an edge in $G^+_x$ and, thus, $v$ and $y$ would be true twins in $G^+_x$.
    This is a contradiction to the previous claim.

  \medskip
  We now show that the constructed tree $T$ is a $k$-leaf root of $G$.
  Notice first that $k \geq 3$, since every graph $G^+_x$ is connected and non-complete.
  Hence, all paths of length $k-2$ used in the construction of $T_h$ have positive length.
  Also, because distinct leaves of $H$ have distinct neighbors, the gluing operations use distinct pendant edges of $T_h$.
  It follows immediately that $T$ is a tree and that its leaf set is exactly $V(G)$.

  Let $v$ and $w$ be two distinct vertices of $G$.
  We distinguish their positions with respect to the graphs $G_x$ and $H$ and prove their correct distance in $T$.
  First, assume that $v, w \in V(G_x) + y$ for a leaf $x$ of $H$ with neighbor $y$.
  The unique $v,w$-path of $T$ is contained in $T_x$, and, therefore,
  $
    \dist_T(v,w)=\dist_{T_x}(v,w).
  $
  Since $T_x$ is a $k$-leaf root of $G_x+y = G[V(G_x) + y]$, we have $vw \in E(G)$ if and only if $\dist_T(v,w) \leq k$.

  Next, let $v,w \in V(h)$.
  By construction of $T_h$, we have
  \[
    \dist_T(v,w) = \dist_{T_h}(v,w) = 2 + (k-2) \dist_H(v,w).
  \]
  If $vw \in E(H)$ then $\dist_H(v,w) = 1$ and, hence, $\dist_T(v,w) = k$.
  Otherwise, we get
  \[
    \dist_H(v,w) \geq 2 \text{ and } \dist_T(v,w) \geq 2+2(k-2) = 2k-2 > k.
  \]
  
  Now let $v \in V(G_x)$ for a leaf $x$ of $H$ with neighbor $y$ and let $w \in V(h) - y$.
  Because $w \not= y$, we have $vw \notin E(G)$ and expect $\dist_T(v,w) > k$.
  Put \mbox{$d = \dist_H(y,w) \geq 1$}.
  The pendant edges of leaf $y$ in $T_x$ and $T_h$ have been identified and, hence,
  \begin{align*}
    \dist_T(v,w) &= \dist_{T_x}(v,y) + \dist_{T_h}(y,w) - 2 = \dist_{T_x}(v,y) + (2 + (k-2) \dist_H(y,w)) - 2\\
    &= \dist_{T_x}(v,y) + (k-2)d.
  \end{align*}
  So, by the observation above,
  \[
    \dist_T(v,w) \geq 3 + (k-2) = k+1.
  \]
  Moreover, every vertex of $G_x$ is adjacent to $y$, and this adjacency is already represented inside $T_x$.

  It remains to consider vertices $v \in V(G_x)$ and $w \in V(G_{x'})$ for two distinct leaves $x$ and $x'$ of $H$ with respective neighbors $y$ and $y'$.
  Since $G_x$ and $G_{x'}$ are independent in $G$, we have $vw \notin E(G)$.
  By precondition, $y \neq y'$ and we let $d = \dist_H(y, y') \geq 1$.
  According to the construction of $T$, we get
  \begin{align*}
    \dist_T(v,w) &= \left(\dist_{T_x}(v,y)-1 \right) + \left(\dist_{T_h}(y, y') - 2\right) + \left( \dist_{T_{x'}}(y',w)-1\right)\\
    &= \dist_{T_x}(v,y) + d(k-2) + \dist_{T_{x'}}(y',w) - 2\\
    &\geq 3 + (k-2) + 3 - 2 = k+2.
  \end{align*}
  We have now covered all possible pairs of vertices and conclude that $T$ is a $k$-leaf root of $G$.
  Next follows the proof of optimality.
  \begin{claim}
    For every $k'$-leaf root $R$ of $G$ with $\pi(k') = p$, it is true $k \leq k'$.
  \end{claim}
    Fix any leaf $x$ of $H$, let $y$ be its neighbor, and choose $q \in N_H(y) - x$ as above.
    We have seen that
    $
      G[V(G_x) + \{y,q\}]
    $
    is isomorphic to $G^+_x$.
    Let $R_x$ be the smallest subtree of $R$ with leaf set $V(G_x) + \{y,q\}$.
    Then $R_x$ is a $k'$-leaf root of $G[V(G_x) + \{y,q\}]$ and, thus, a $k'$-leaf root of $G^+_x$ up to isomorphism.
    Since $k_x$ is minimum among all leaf-root parameters for $G^+_x$ with parity $p$, we obtain $k_x \leq k'$.
    This holds for every leaf $x$ of $H$ and, consequently,
    \[
      k = \max\{k_x\mid x\text{ is a leaf of }H\} \leq k'.
    \]
    Hence, $T$ is optimal among all leaf roots of $G$ whose parameter has parity $p$ and the claim's proof is complete.
  
  \medskip
  It remains to consider the running time.
  The decomposition of $G$ into $H$ and the graphs $G_x$ can be done via the block-cut-tree $\text{BCT}$ of $G$, which can be computed in $\cO(n+m)$ time~\cite{Tarjan72}.
  In fact, the graphs $G^+_x$ can altogether be represented in $\cO(n+m)$ time.
  Indeed, the graphs $G_x$ are pairwise disjoint, and
  $
    \sum_{x\in L} |V(G^+_x)| = \sum_{x\in L} \left(V(G_x)+2\right) = \cO(n)
  $.
  Moreover, $E(G^+_x)| = |E(G_x)| + |V(G_x)| + 1$ and, consequently,
  $
    \sum_{x \in L} |E(G^+_x)| = \cO(n+m)
  $.
  The linear-time leaf-root algorithm for chordal cographs therefore computes all pairs $(T'_x,k_x)$ in $\cO(n+m)$ time altogether~\cite{LeR24}.
  The value $k$ is found in $\cO(n)$ time by a simple $\max$-search.
  According to~\cite{LeR24}, all extensions
  $
    T''_x=\eta(T'_x,(k-k_x)/2)
  $
  can be carried out in $\cO(n)$ time altogether when the trees are represented in compressed form.
  Cutting off the artificial leaves and obtaining the trees $T_x$ takes time linear in the total size of these compressed trees and, hence, $\cO(n)$ time.
  The compressed tree $T_h$ has $\cO(|V(H)|)$ vertices and weighted edges and is constructed in $\cO(n)$ time.
  Finally, every gluing operation requires only constant time, including splitting off the last unit edge of a compressed pendant edge, and there are at most $n$ leaves in $H$.
  Altogether, a compressed representation of $T$ is computed in $\cO(n+m)$ time.
  This proves the lemma.
\qed\end{proof}

\section*{Detailed Disclosure on the Use of Generative AI}
\label{sec:generative_ai}

During the preparation of this paper, the authors used OpenAI's GPT-5.6 Sol as a generative-AI assistant.
All principal mathematical ideas and results, except Lemma~\ref{thm:bull-free_prp_int_path_power}, originated independently with the authors.

AI assistance was used to critically evaluate all proofs, leading to the identification and correction of several minor proof gaps and the adoption of sharper local arguments.
Lemma~\ref{thm:bull-free_prp_int_path_power} and substantial parts of its proof originated from an AI-assisted examination of Condition~(i) in Lemma~\ref{lem:bull_dart-free_structure}.
This examination revealed a substantially stronger structural restriction than initially anticipated, ultimately leading to the path-power characterization.

AI also contributed substantially to the development of the technical proof of Lemma~\ref{lem:bull_gem-free_structure:leaf_root}, particularly its case analysis.
The underlying optimality criterion and its justification, however, were developed independently by the authors.

Further AI assistance included language and presentation refinement, particularly in the introduction and preliminaries, as well as literature searches and bibliographic research.
All AI-assisted material was critically reviewed and verified by the authors, who take full responsibility for the contents of this paper.

\end{document}